\newcommand*{\ATLASLATEXPATH}{}
\documentclass[cernpreprint,texlive=2011,txfonts,UKenglish,texmf]{atlasdoc}
\pdfoutput=1

\usepackage[biblatex=false, siunitx=false]{\ATLASLATEXPATH atlaspackage}
\usepackage{\ATLASLATEXPATH atlascontribute}
\usepackage{\ATLASLATEXPATH atlasphysics}

\usepackage{units}
\usepackage[hyphens]{url}
\usepackage{amsmath}
\usepackage{graphicx}
\usepackage{amssymb}
\usepackage{lineno}
\usepackage{multirow}
\usepackage{a4wide}

\graphicspath{{logos/}{figures/}}

\usepackage{mydocument-defs}

\hypersetup{pdftitle={ATLAS draft},pdfauthor={The ATLAS Collaboration}}

\AtlasTitle{Search for heavy long-lived multi-charged particles in $pp$ collisions at $\sqrt{s}=8$~\TeV{} using the ATLAS detector}

\author{The ATLAS Collaboration}

\PreprintIdNumber{CERN-PH-EP-2015-049}

 \AtlasJournal{\EPJC}

\AtlasAbstract{%
A search for heavy long-lived multi-charged particles is performed 
using the ATLAS detector at the LHC. 
Data collected in 2012 at $\sqrt{s}=8$~\TeV{} from $pp$ collisions 
corresponding to an integrated luminosity of \intlumi{} are examined.
Particles producing anomalously 
high ionisation, consistent with long-lived massive particles with electric charges from $|q|=2e$ 
to $|q|=6e$ are searched for. No signal candidate events are observed, and 95\% confidence level 
cross-section upper limits are interpreted as lower mass limits for a 
Drell--Yan production model. 
The mass limits range between $660$ and \unit[785]{\GeV}.
}

\begin{document}

\maketitle

\tableofcontents

\section{Introduction}
\label{sec:intro}

This article describes a search for heavy long-lived\footnote{The term long-lived 
in this paper refers to a particle that does not decay within 
the ATLAS detector and penetrates its full depth.} 
multi-charged particles (MCPs) in $\sqrt{s}=$ \unit[8]{\TeV} $pp$ collisions data 
collected in 2012 by the ATLAS detector at the CERN 
Large Hadron Collider (LHC). Data taken in stable beam conditions and with all 
ATLAS subsystems operational are used, resulting in an 
integrated luminosity of \intlumi{}.
The search is performed in the MCP mass range of \unit[50--1000]{\GeV}, for 
electric charges\footnote{Wherever a charge is quoted for the
exotic particles, the charge conjugate state is also
implied.} ${|q|=ze}$, with the charge numbers ${z=2}$, $3$, $4$, $5$, and $6$.
The observation of such particles possessing an electric charge above the elementary 
charge $e$ would be a signature for physics beyond the Standard Model. 
Several theories predict such particles, including the 
almost-commutative model~\cite{Stephan:2005uj}, the walking 
technicolor model~\cite{Sannino:2004qp}, and the 
left-right symmetric model~\cite{Mohapatra:1974hk}, which predicts a doubly charged Higgs boson.
Any observation of the particles predicted by the first two models could have 
implications for the formation of composite dark matter:  
the doubly charged particles (or, in general, particles with an even charge $|q| = 2ne$) 
could explain many results of experimental searches for dark matter~\cite{Khlopov:2011tn}.
No such particles have been observed so far in cosmic 
ray~\cite{SLIM} or collider searches, including several recent searches 
at the Tevatron~\cite{Acosta:2005np} and the 
LHC~\cite{HIP2010,Qballs2011,QballsCms20112012}.

MCPs are highly ionising, and thus leave an abnormally 
large ionisation signal, \dedx{}.
A search for such particles traversing the ATLAS detector
leaving a track in the inner tracking detector, and producing a 
signal in the muon spectrometer, is reported.
A purely electromagnetic coupling, 
proportional to the electric charge of the MCPs, 
is assumed 
for the production model.  
In this model, MCPs are produced in pairs via the Drell--Yan (DY) process 
with only photon exchange included.

This analysis is also sensitive to fractionally charged ($z>1$, non-integer) particles, 
but has not been interpreted explicitly for such charges.
The signal efficiency in a search for MCPs with charge numbers 
higher than $z=6$ is expected to be less than $5\%$ due to the signal 
particle's low velocity. 
Such low efficiencies require a different approach, and corresponding model 
interpretations are not covered in this paper.

\section{The ATLAS detector}
\label{sec:ATLAS}

The ATLAS detector~\cite{atlas:2008} covers nearly the entire
solid angle around the collision point. It consists of an inner
tracking detector (ID) comprising a silicon pixel detector (pixel), a silicon
microstrip detector (SCT) and a transition radiation tracker
(TRT). The pixel detector typically provides one precise space-point 
measurement per track from each of its three layers. The SCT consists of four 
times two layers of silicon sensors arranged with small stereo angle, 
typically providing eight measurements per track.
The TRT, covering the pseudorapidity range $|\eta|<2.0$,\footnote{ATLAS uses a right-handed 
coordinate system with its origin at the 
nominal interaction point (IP) in the centre of the detector and the $z$-axis 
along the beam pipe. The $x$-axis points from the IP to the centre of the LHC ring, 
and the $y$-axis points upward. Cylindrical coordinates $(r,\phi)$ are used in the 
transverse plane, $\phi$ being the azimuthal angle around the $z$-axis. The 
pseudorapidity is defined in terms of the polar 
angle $\theta$ as $\eta=-\ln\tan(\theta/2)$. Angular distance is measured in units 
of $\Delta R \equiv \sqrt{(\Delta\eta)^{2} + (\Delta\phi)^{2}}$.} is a straw-based 
tracking detector capable of particle identification via
transition radiation and ionisation energy loss measurements~\cite{TRTPID}. 
A typical track crosses $32$ straws. Discriminators are used to compare the 
signal from a straw with low and high thresholds (HT) using the TRT front-end electronics. 
The HT is designed to discriminate between energy depositions from transition radiation 
photons and the energy loss of minimum ionising particles. Roughly three times 
the energy deposition of a minimum ionising particle is needed for a HT hit. 
MCPs would produce a large number of HT hits along their trajectories due to their high 
level of ionisation.

The ID is surrounded by a thin superconducting solenoid providing a \unit[2]{T}
axial magnetic field, and by high-granularity lead--liquid argon 
(LAr) sampling electromagnetic calorimeters.
An iron--scintillator tile calorimeter
provides hadronic energy measurements in the central pseudorapidity region.
The endcap and forward regions are instrumented with
LAr calorimeters for electromagnetic and hadronic energy
measurements. In this analysis, the calorimeters are used only as passive absorbers. 
The calorimeter system is surrounded
by a muon spectrometer (MS) incorporating three superconducting toroidal
magnet assemblies. The MS is instrumented with tracking detectors designed 
to measure the momenta of muons that traverse the ATLAS calorimeters.
The resistive-plate chambers (RPC) in the barrel region (${|\eta| < 1.05}$) and
the thin-gap chambers (TGC) in the endcaps regions (${1.05 < |\eta| < 2.4}$) provide signals for the trigger.
Monitored drift tube (MDT) chambers provide typically 20--25 hits per crossing 
track in the pseudorapidity range ${|\eta| < 2.7}$, from which a high precision 
momentum measurement is derived.

The amount of material in the ID varies from one-half to two radiation 
lengths. The overall amount of material traversed by the MCP, which includes the calorimeters 
and the MS, may be as high as $75$ radiation lengths. Muons typically 
lose \unit[3]{\GeV} penetrating the calorimeter system. 
The energy loss for MCPs with charge ${|q|=ze}$ would be $z^{2}$ times this value, i.e. up to \unit[110]{\GeV} 
for ${z=6}$.

All momentum values quoted in this paper are measured by the MS, 
after the energy loss in the calorimeters. 
Charged-particle trajectories are reconstructed using standard algorithms. 
Since these assume particles have $z=1$, the momenta of MCPs are 
underestimated by a factor $z$, as the track curvature is proportional to ${\pT / z}$.

\section{Simulated Monte Carlo samples}
\label{sec:signalsimulation}

Benchmark samples of simulated events with MCPs
are generated 
for a mass of \unit[50]{\GeV} and for a range of masses between $100$ and \unit[1000]{\GeV} 
in steps of \unit[100]{\GeV}, with charges 
$ze$, $z=2$, $3$, $4$, $5$, and $6$. 
Pairs of MCPs are generated via the lowest-order DY process implemented in
{\sc MadGraph5}~\cite{madgraph5}. The DY production process models the 
kinematic distributions and determines the cross-sections used for limit setting. 
Typical values for the cross-sections 
range from hundreds of picobarns for a mass of \unit[50]{\GeV} down 
to a hundredth of a femtobarn for a mass of \unit[1000]{\GeV} 
(Fig.~\ref{fig:limtrack}).
Events are generated using the {\sc CTEQ6L1}~\cite{Pumplin:2002vw} parton 
distribution functions, and {\sc Pythia} version 8.170~\cite{Sjostrand:2006za,Pythia8} is
used for hadronisation and underlying-event generation. 
Simulated samples with muons from $Z\to\mu\mu$ decays are 
generated using {\sc Pythia} version 8.170 and the 
{\sc CT10}~\cite{Lai:2010vv} parton distribution functions with the {\sc AU2} tune~\cite{AU2Tune}. 
A {\sc Geant4} simulation~\cite{Agostinelli:2002hh, Aad:2010ah}
 is used to model the response of the ATLAS detector. 
Each simulated hard scattering event is overlaid with simulated minimum bias 
events (``pile-up'') generated with {\sc Pythia} in order to reproduce the observed 
distribution of the number of proton--proton collisions per bunch crossing.
The simulated events 
are reconstructed and analysed in the same way as the experimental data.

\section{Candidate and event selection}
\label{sec:selection}

Because the MCPs in this search are assumed to be long-lived and 
therefore traverse the entire ATLAS detector, candidates 
are initially selected with the MS.
The search, which is restricted to the ${|\eta| < 2.0}$ pseudorapidity range, 
is based on an analysis of specific ionisation losses in several 
sub-detector systems and of the fraction of TRT straws on the track with a signal 
amplitude exceeding the HT. 
The search is logically divided into four steps: trigger and event selection, 
preselection, tight selection and final selection. 
The tight and final selection steps rely on the ionisation estimators, 
which are introduced in the following section.
An event is considered to be a candidate event
if it has at least one candidate MCP (a reconstructed particle, which satisfies 
all selection criteria).

\subsection{Ionisation estimators}

The average specific energy loss, \dedx{}, is described by the Bethe--Bloch
formula~\cite{bethe}. Since a particle's energy loss increases quadratically with its charge, 
an MCP would leave a very
characteristic signature of high ionisation in the detector. Estimates of
\dedx{} are evaluated for the pixel, TRT and MDT sub-detector systems.
All three quantities are based on an underlying measurement of time-over-threshold:
the time interval where a signal amplitude exceeds a certain threshold 
is correlated with the deposited energy.

The significance of the \dedx{} variable in each sub-detector is defined by comparing the observed signal,
$\text{\dedx{}}_{\text{track}}$, with that expected from a highly relativistic muon:

\begin{equation}
S(\dedx{}) = \frac{\text{\dedx{}}_{\text{track}} - \langle\text{\dedx{}}_{\mu}\rangle}{\sigma(\text{\dedx{}}_{\mu})}.
\end{equation}
Here $\langle\text{\dedx{}}_{\mu}\rangle$ and $\sigma(\text{\dedx{}}_{\mu})$ 
represent, respectively, the mean and the root-mean-square width of the \dedx{} distribution for such muons in data.
For this procedure, a control sample of muons was obtained from $Z\to\mu\mu$ events. 
Each muon was required to be matched to a good-quality track in the ID with
${\pT > \unit[24]{\GeV{}}}$ and ${|\eta| < 2.0}$, be 
isolated, i.e.\ to carry at least $90\%$ of the total \pT within the surrounding $\Delta R<0.2$ cone, 
and belong to an oppositely charged pair with dimuon mass 
between \unit[81]{\GeV{}} and \unit[101]{\GeV{}}. These requirements effectively 
suppress muons from other processes reducing such backgrounds to a negligible level.

In addition to the \dedx{} estimates, the fraction 
of TRT hits passing the high threshold, $f^{\textrm{HT}}$, is another estimator of energy loss. 

\begin{figure}[h!]
  \begin{center}
    \includegraphics[width=0.45\textwidth]{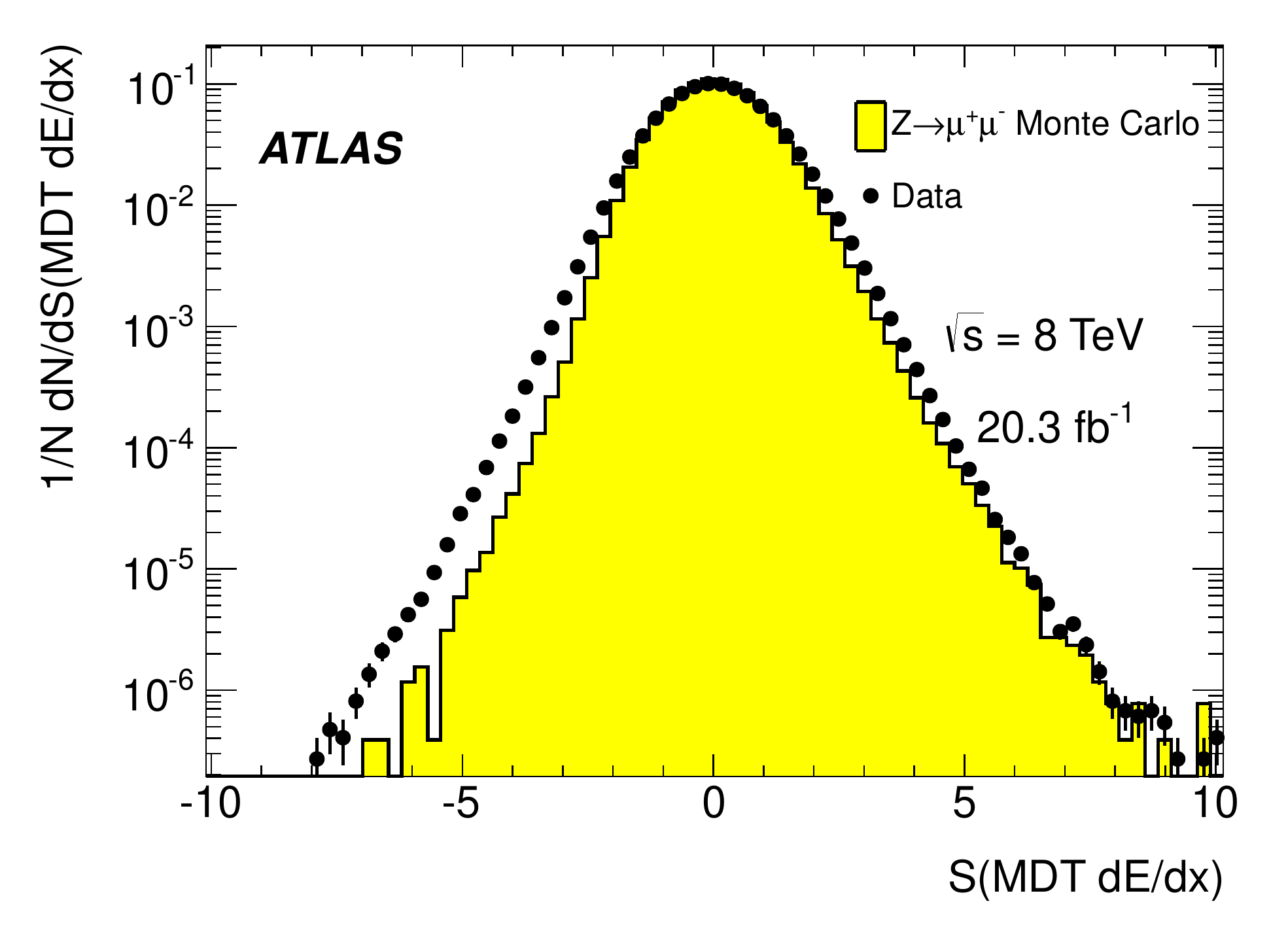}
    \includegraphics[width=0.45\textwidth]{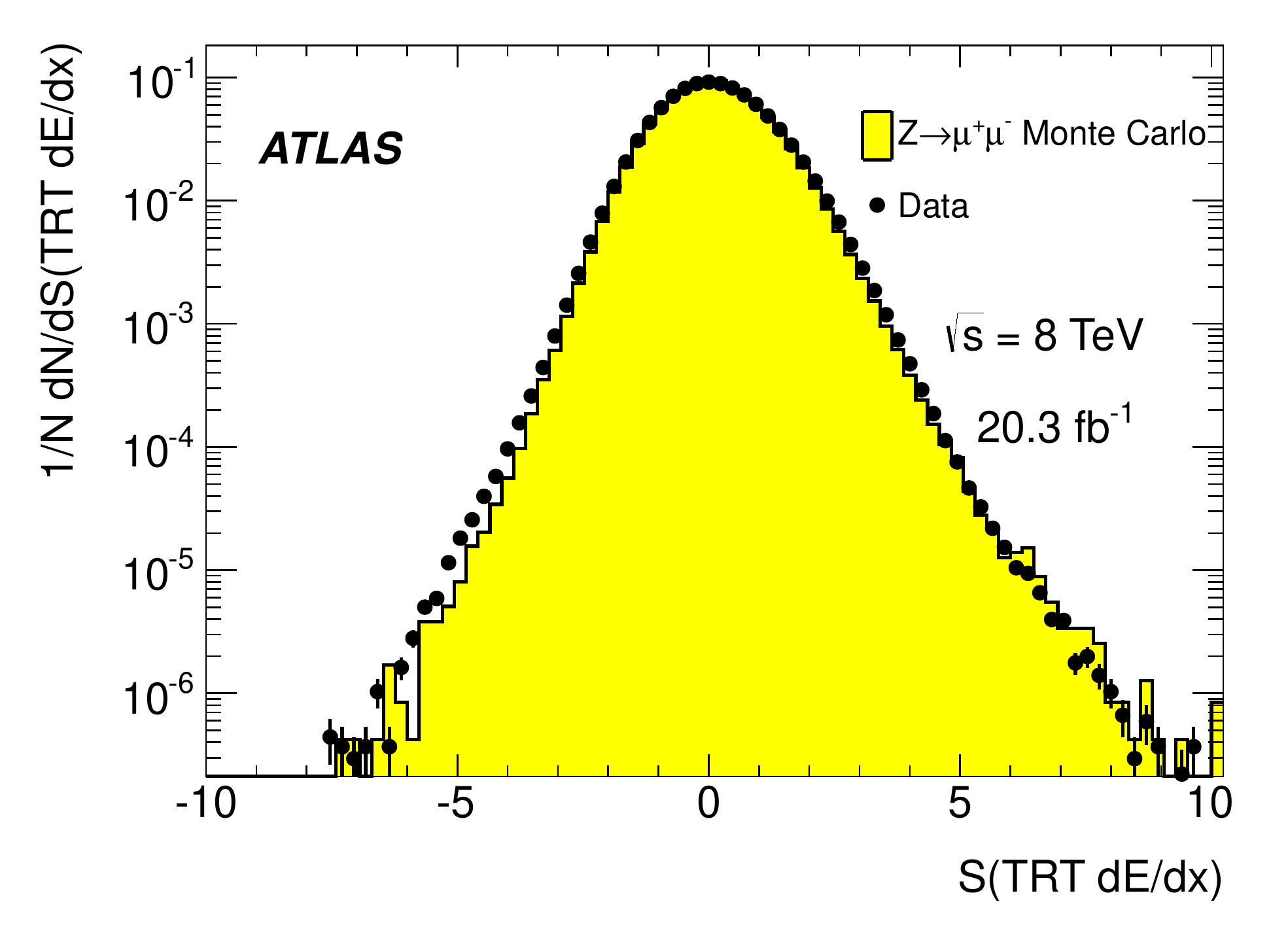}
  \end{center}
  \caption{Normalised distributions of the \dedx{} significance in the MDT, $S$(MDT \dedx{}), (left) 
    and in the TRT, $S$(TRT \dedx{}), (right) for muons from $Z\rightarrow \mu\mu$ events in data and simulation.}
  \label{fig:dataMCagreement_MdtAndTrt}
\end{figure}

\begin{figure}[h!]
  \begin{center}
    \includegraphics[width=0.45\textwidth]{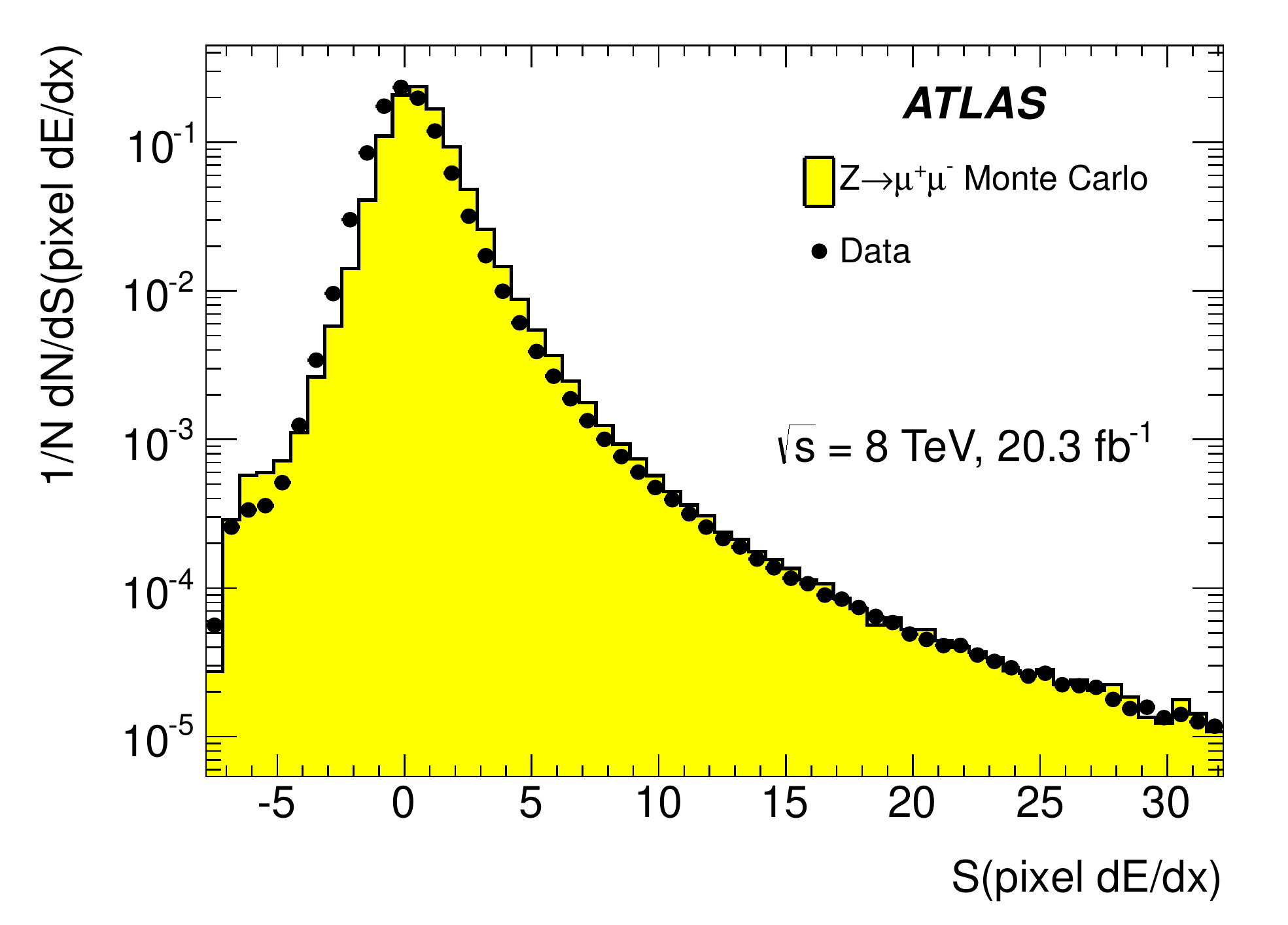}
    \includegraphics[width=0.45\textwidth]{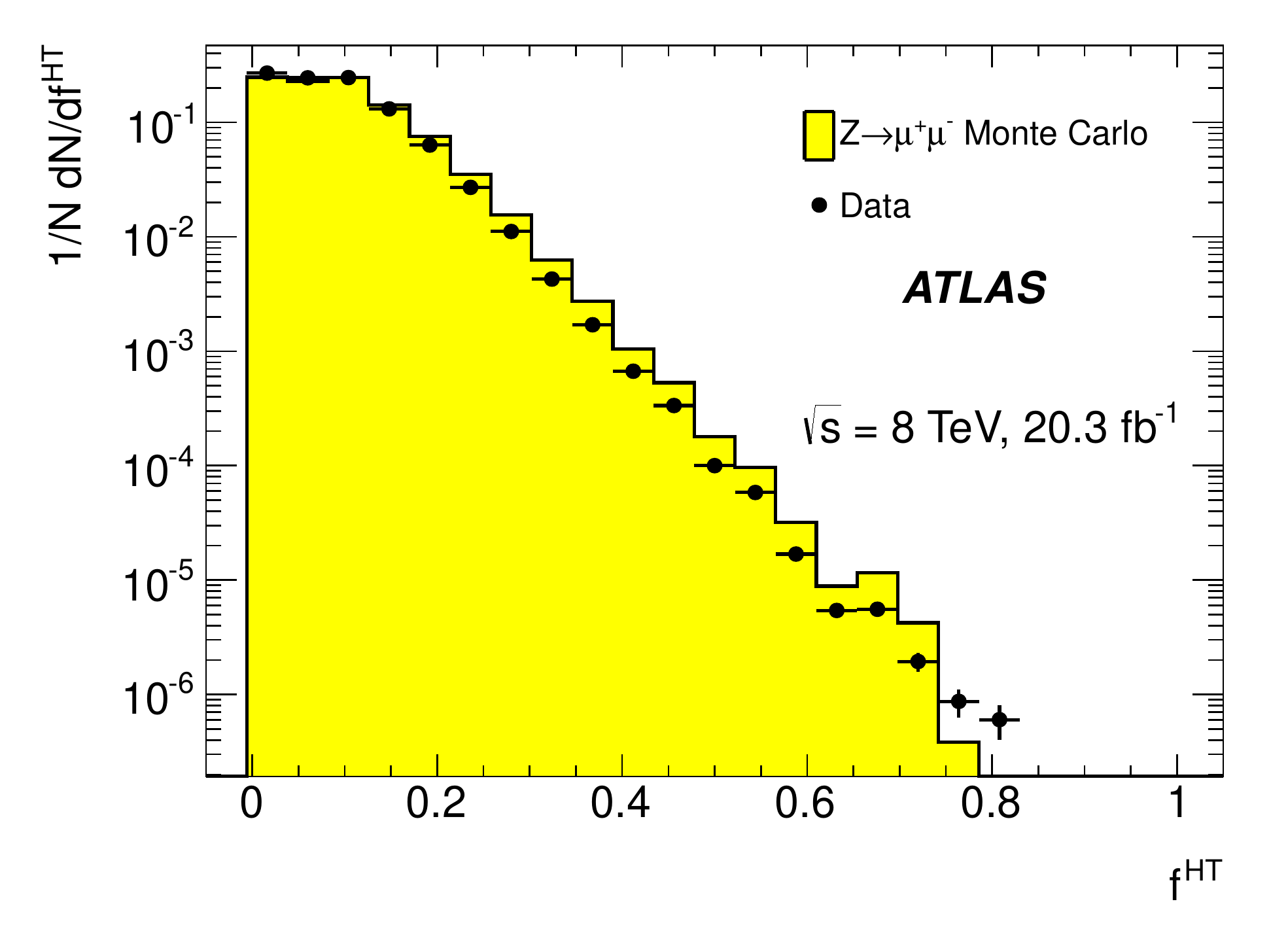}
  \end{center}
  \caption{Normalised distributions of the \dedx{} significance in the pixel system, $S$(pixel \dedx{}), (left) 
    and $f^{\textrm{HT}}$, the fraction of TRT hits passing the high threshold, (right)
    for muons from $Z\rightarrow \mu\mu$ events in data and simulation.}
  \label{fig:dataMCagreement_PixelAndfHT}
\end{figure}

In order to investigate whether the relevant variables are modelled properly, 
muons from $Z\rightarrow \mu\mu$ decays are compared between data and simulation. 
Figure~\ref{fig:dataMCagreement_MdtAndTrt} shows the comparison for the MDT and TRT \dedx{} significances, 
and Fig.~\ref{fig:dataMCagreement_PixelAndfHT} for the pixel \dedx{} significance and $f^{\textrm{HT}}$.
In general, Figs.~\ref{fig:dataMCagreement_MdtAndTrt} and~\ref{fig:dataMCagreement_PixelAndfHT} 
demonstrate good agreement between simulated and experimental data for the four 
selection variables. This is especially true on the high side of the distributions, 
which is most relevant for the analysis. The small differences 
observed, particularly for the $S$(MDT \dedx{}) 
variable, have only minor effects on the analysis, and are accounted for 
as systematic uncertainties, described in Sect.~\ref{sec:systematics}.
The behaviour of all four selection variables 
is found to be stable with respect to $\eta$, $\phi$ and \pT.

Detailed studies of energy loss vs.\ momentum distributions were 
performed for the pixel~\cite{ATLAS:2011gea} and TRT~\cite{TRTPID} 
detectors, as well as for the relativistic rise domain of the Bethe--Bloch formula 
in the MDT. These results assure that the moderate ionisation 
levels (like for ${z=2}$ particles) are correctly described in the simulated data. The responses to 
the higher charge particles are well above the selection requirements 
(conservatively defined for the ${z=2}$ particles), and so the analysis is not 
sensitive to the precise mean position of the distributions, which may be shifted by any 
potential saturation effects.

\subsection{Trigger and event selection}
\label{sec:TriggAndEvSel}

Events collected with a single-muon trigger~\cite{Aad:2014sca}
with a transverse momentum threshold 
of \linebreak ${\pT/z = \unit[36]{\GeV{}}}$ are considered.
This trigger is only sensitive to particles with velocity $\beta=v/c>0.6$ 
due to a timing window, in which particles should reach the MS.
To compensate for inefficiencies in the single-muon trigger, 
an additional calorimeter-based trigger 
with a missing transverse momentum (\etmiss{}) threshold of \unit[80]{\GeV{}} is employed.
Particles reconstructed in the MS are not accounted for in the trigger \etmiss{} calculation,
thus they contribute to the missing transverse momentum value directly. 
Large missing transverse momentum can also be due to an asymmetry between the energy depositions 
in calorimeters of the two MCPs. 
In case an event is selected by both of these triggers, 
it is assigned to the single-muon trigger for the following analysis.
The \etmiss{} trigger recovers up to $10\%$ of events missed by the 
single-muon trigger.
 
Events are further required 
to contain at least one muon candidate with either
${\pT/z > \unit[75]{\GeV{}}}$ (single-muon trigger) or 
with ${\pT/z > \unit[60]{\GeV{}}}$ (\etmiss{} trigger).\footnote{Information
on the MDT and TRT \dedx{} is not available in the
standard ATLAS data stream. Hence, this analysis is based on special
streams which include this information. The $\pT$ requirements for muons
given here are imposed in the preparation of these streams and are not optimised for the
current analysis.}

\subsection{Candidate track preselection}
\label{sec:preselection}

Each candidate track reconstructed in the MS with at least $7$ MDT hits should match a high-quality 
track in the ID. Such an ID track is required to have at least $6$ SCT hits and $10$ TRT hits, 
and to originate less than \unit[1.5]{mm} in both the longitudinal ($|z_{0}\sin\theta|$) and 
transverse ($|d_{0}|$) directions from the primary interaction point, determined via standard technique 
as described in Ref.~\cite{Piacquadio:2008zzb}. Each candidate track must also be within the acceptance 
region of the TRT (${|\eta| < 2.0}$), have ${\pT/z > \unit[40]{\GeV{}}}$ for 
events collected with the single-muon trigger 
or ${\pT/z > \unit[30]{\GeV{}}}$ for those collected with the \etmiss{} trigger. 
The efficiency of the ID track reconstruction varies between $96\%$ and $98\%$ 
for all MCP charge values considered.

In order to reduce the background of high ionisation signals from two or more 
tracks firing the same TRT straws or MDT tubes, each candidate is required 
not to have an adjacent track with ${\pT/z > \unit[5]{\GeV{}}}$ within ${\dR < 0.01}$.

The preselected data sample (selected with these requirements) is 
completely dominated by muons, even in the presence of a possible signal.

\subsection{Tight selection}
\label{sec:tight_selection}

\begin{figure}[h!]
  \begin{center}
    \includegraphics[width=0.45\textwidth]{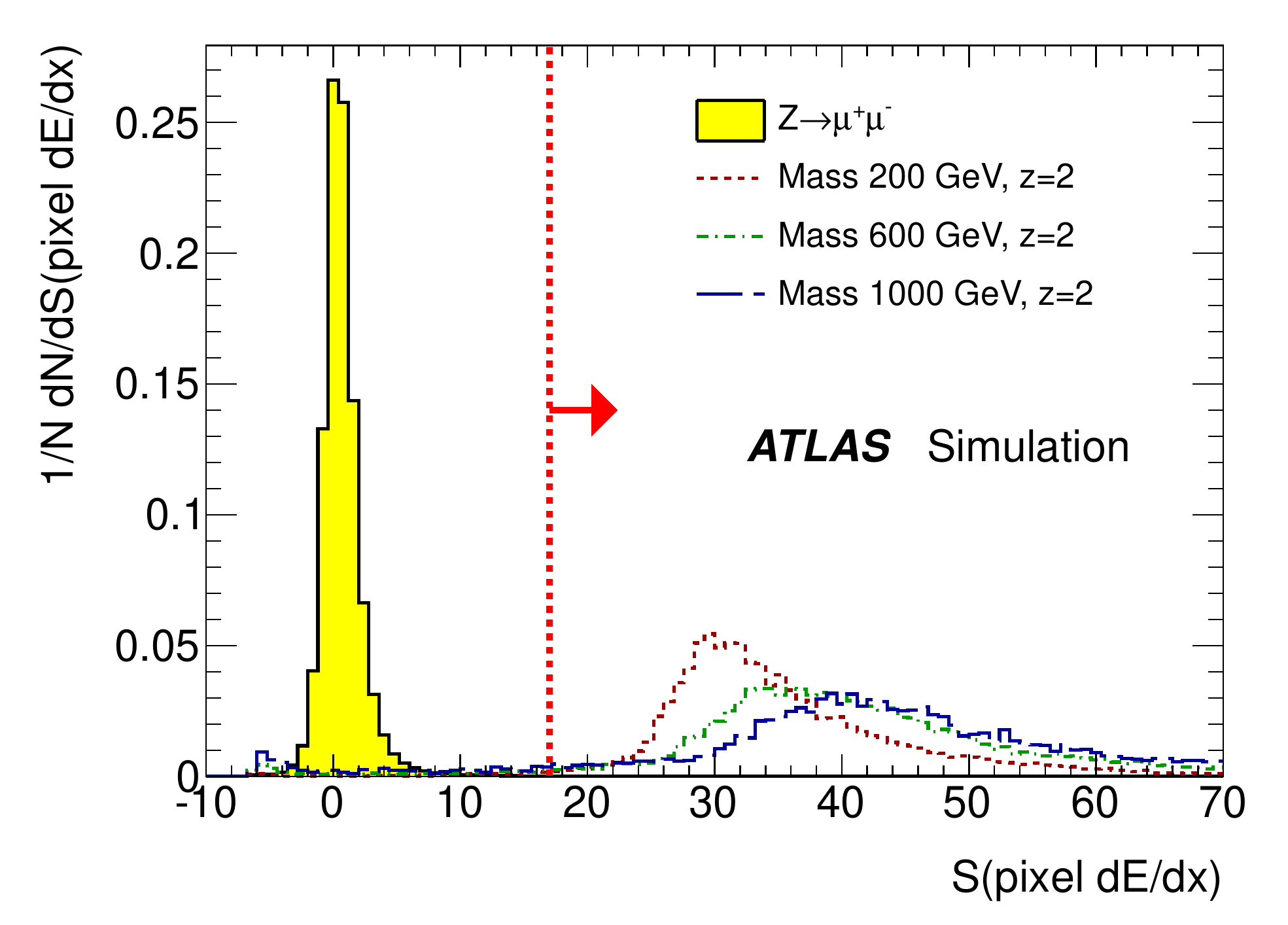}
    \includegraphics[width=0.45\textwidth]{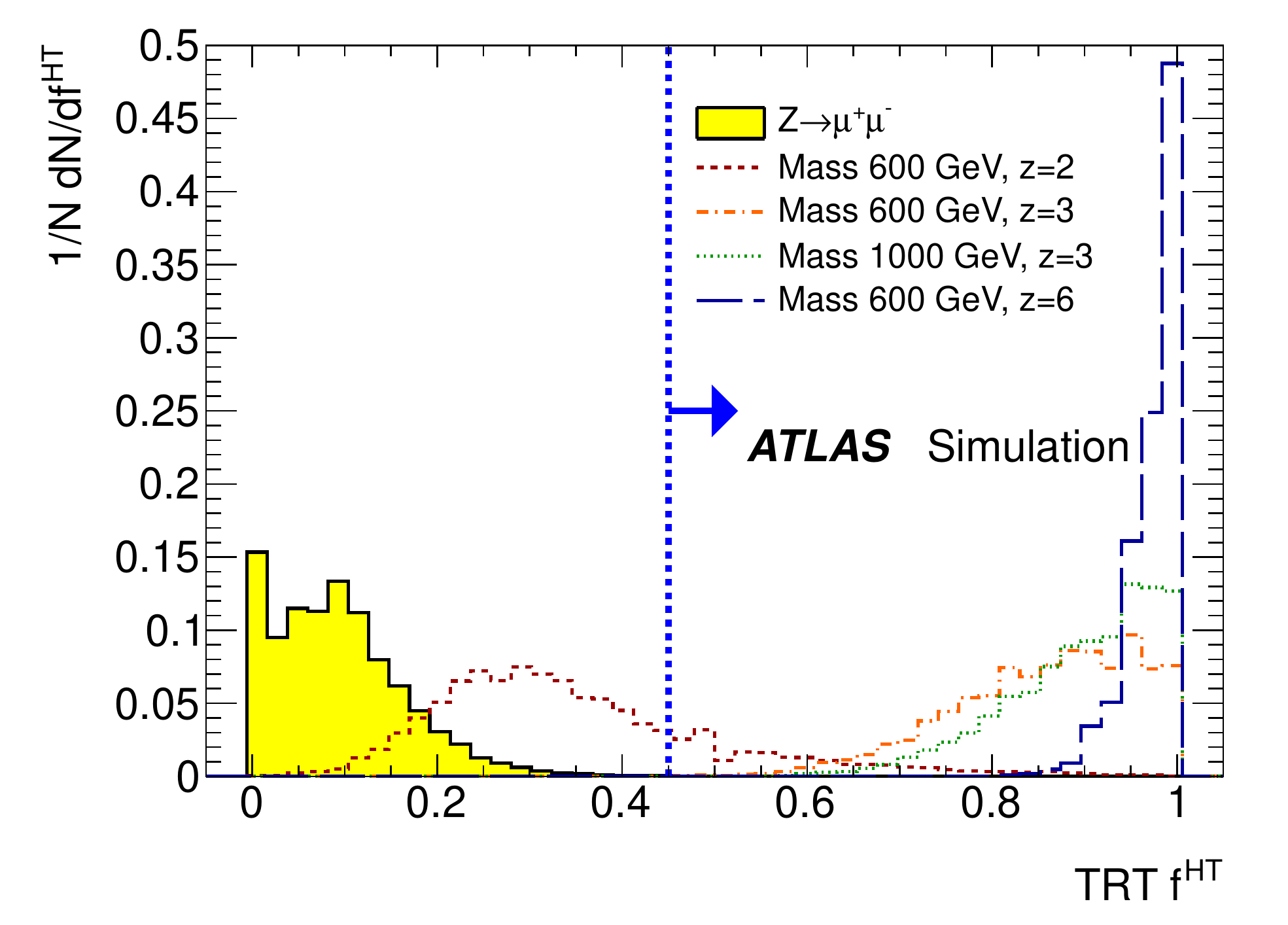}
  \end{center}
  \caption{Normalised distributions of the \dedx{} significance in the pixel system, 
    $S$(pixel \dedx{}), (left) and $f^{\textrm{HT}}$ (right)
    for simulated muons from $Z\rightarrow \mu\mu$ events and MCPs passing the preselection requirements.
    Signal distributions are shown for ${z=2}$ and masses of $200$, $600$ and \unit[1000]{\GeV{}} (left)
    and for ${z = 3}$ and $6$ for a mass of \unit[600]{\GeV{}} and, additionally, 
    for ${z = 3}$ and a mass of \unit[1000]{\GeV{}} (right). For comparison, the ${z=2}$ distribution 
    is also shown on the right plot, although $f^{\textrm{HT}}$ is not used in the ${z=2}$ MCPs search.
    The red (blue) dotted line indicates the thresholds of the selection criteria for the ${z = 2}$ (${z \geq 3}$) case.}
  \label{fig:dEdxSignificanceSeparation_Pixel_And_TRTfHT}
\end{figure}

The tight selection of highly ionising candidates is based on 
$S$(pixel \dedx{}) for MCPs with ${z=2}$, and on $f^{\textrm{HT}}$ for MCPs with $z\geq3$.
As seen in Fig.~\ref{fig:dEdxSignificanceSeparation_Pixel_And_TRTfHT}, 
$S$(pixel \dedx{}) is a powerful discriminator for particles with $z=2$. 
The signal region is defined to be the region with significance greater than \pixSCut{}.
For higher values of $z$, the pixel readout saturates and the 
charge information for a particular pixel is lost.
Therefore, to search for particles 
with ${z \geq 3}$, $f^{\textrm{HT}}$ 
(see Fig.~\ref{fig:dEdxSignificanceSeparation_Pixel_And_TRTfHT}) is used as a discriminating
variable instead. The signal region is defined by requiring $f^{\textrm{HT}}$
to be above \htCut{}. 

This tight selection using $S$(pixel \dedx{}) 
or $f^{\textrm{HT}}$ criteria reduces the background contribution (mainly the high-\pT muons) 
by almost three orders of magnitude for both the ${z = 2}$ and ${z \geq 3}$ cases, 
while keeping an efficiency above $95\%$ for the signal.

\subsection{Final selection}
\label{sec:final_selection}

\begin{figure}[h!]
  \begin{center}
    \includegraphics[width=0.45\textwidth]{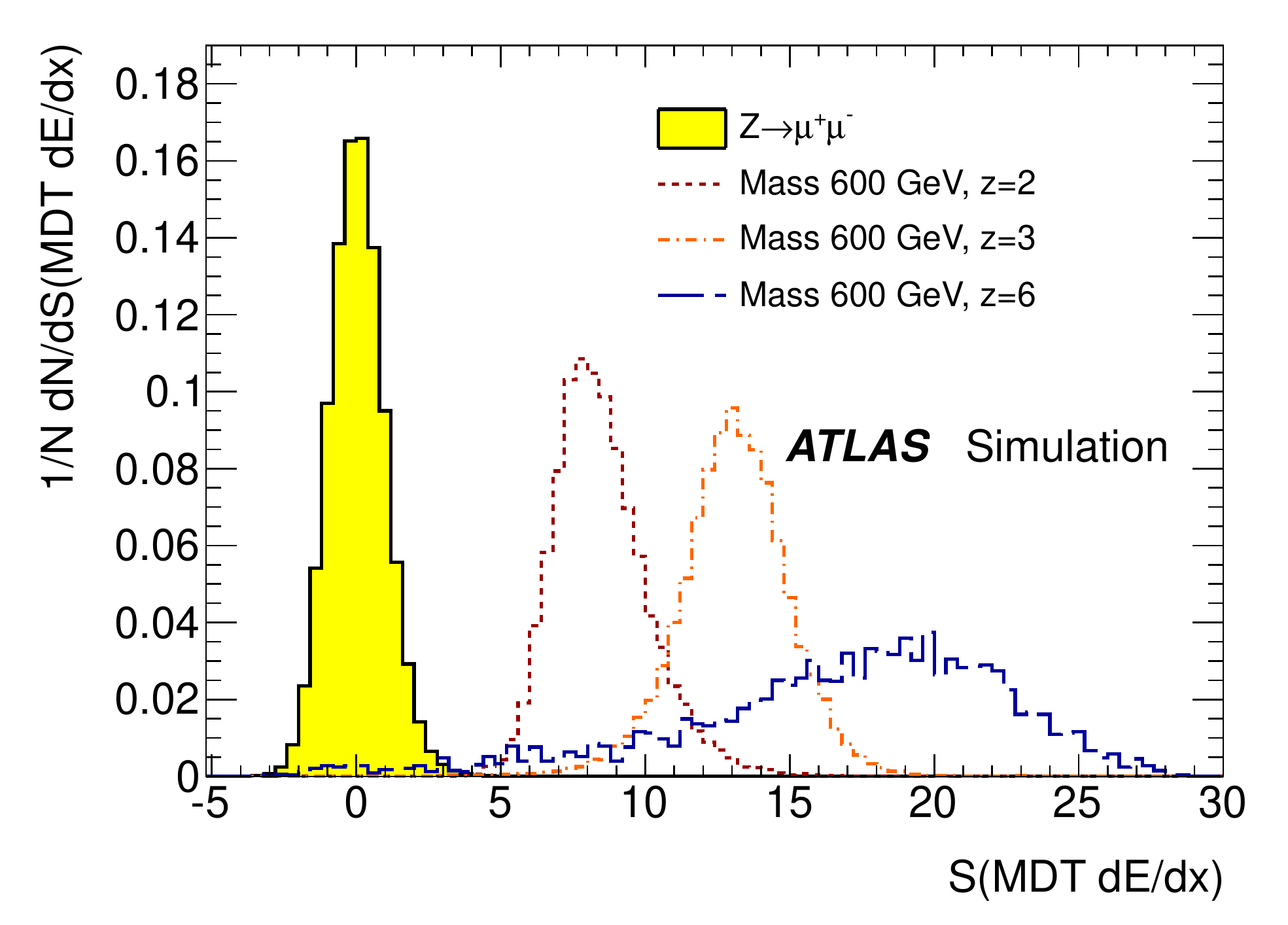}
    \includegraphics[width=0.45\textwidth]{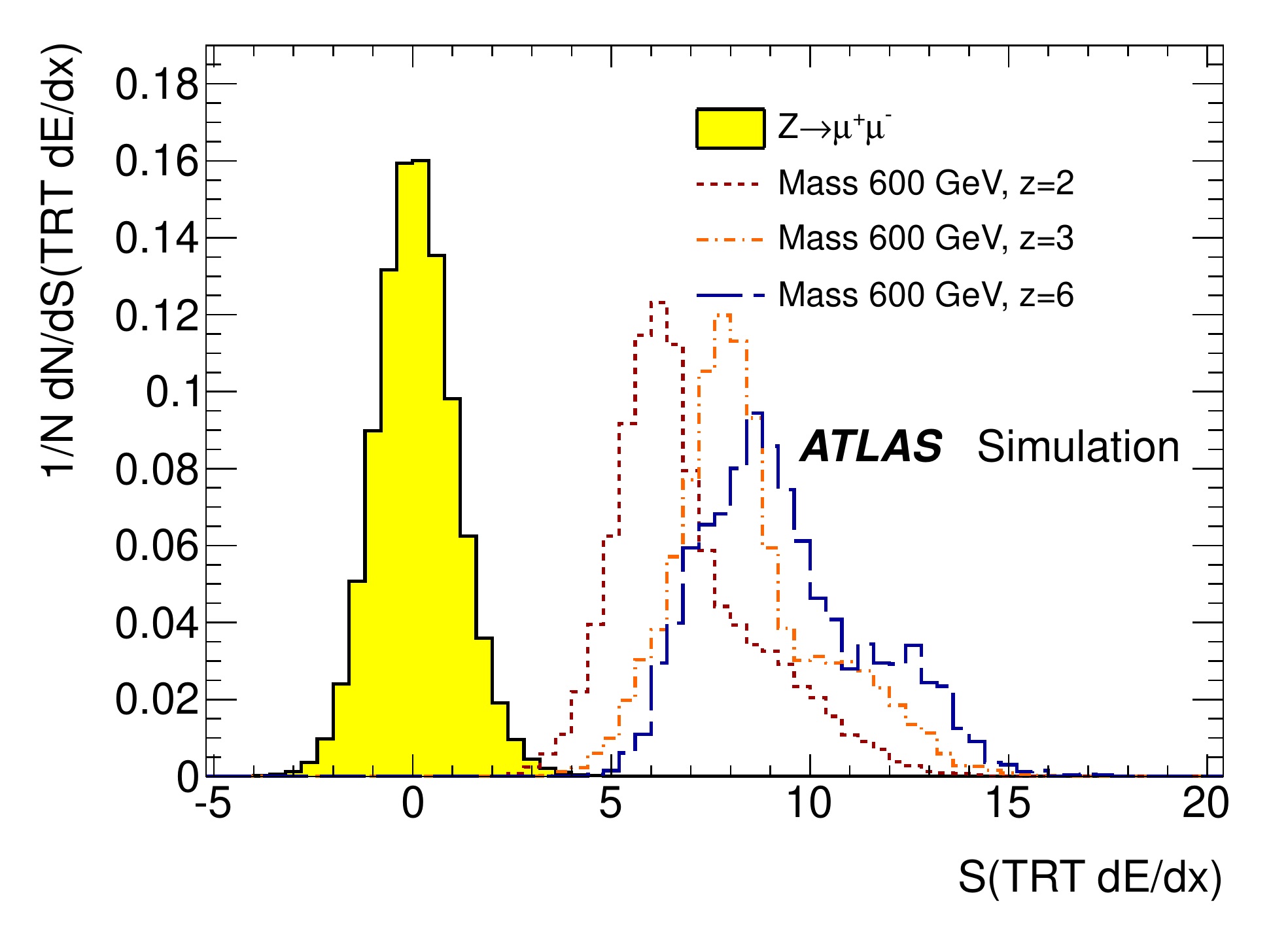}
  \end{center}
  \caption{Normalised distributions of the \dedx{} significance in the 
    MDT, $S$(MDT \dedx{}), (left) and in the TRT, $S$(TRT \dedx{}), (right) 
    for simulated muons from $Z\rightarrow \mu\mu$ events and MCPs.
    Signal distributions are shown for ${z=2}$, $3$ and $6$, for a mass of \unit[600]{\GeV{}}.
}
  \label{fig:dEdxSignificanceSeparation_MdtAndTrt}
\end{figure}

In the final step of the search, $S$(MDT \dedx{}) and $S$(TRT \dedx{}) 
are used as additional discriminating variables to separate the signal and background. 
Figure~\ref{fig:dEdxSignificanceSeparation_MdtAndTrt} shows the distributions of 
these variables for simulated muons from
$Z\to\mu\mu$ production compared to those of signal particles
for different charges (${z=2}$, $3$ and $6$) and for a mass of \unit[600]{\GeV{}}. 
It demonstrates good separation between signal and background, 
which increases with increasing charge. 
The $S$(MDT \dedx{}) distribution shape broadens with charge because of a larger track curvature,
which hinders the track reconstruction algorithms from finding all hits on the track, thus
decreasing the accuracy of the ionisation loss measurement.
The detailed response for 
these higher charge particles may not be perfectly
modelled by the simulation due to saturation effects. However, since these detectors 
do not lose signal at saturation, their
\dedx{} response would certainly be higher than that of ${z = 2}$
particles.

The \dedx{} significance strongly depends on the particle's charge and on its 
velocity (for a given velocity, it does not depend on the particle's mass). For the MCPs under study, 
the variation of velocity ($0.6\leq\beta<1$) leads to a change in \dedx{} significances 
by up to $30\%$.

\begin{figure}[h!]
  \begin{center}
    \includegraphics[width=0.45\textwidth]{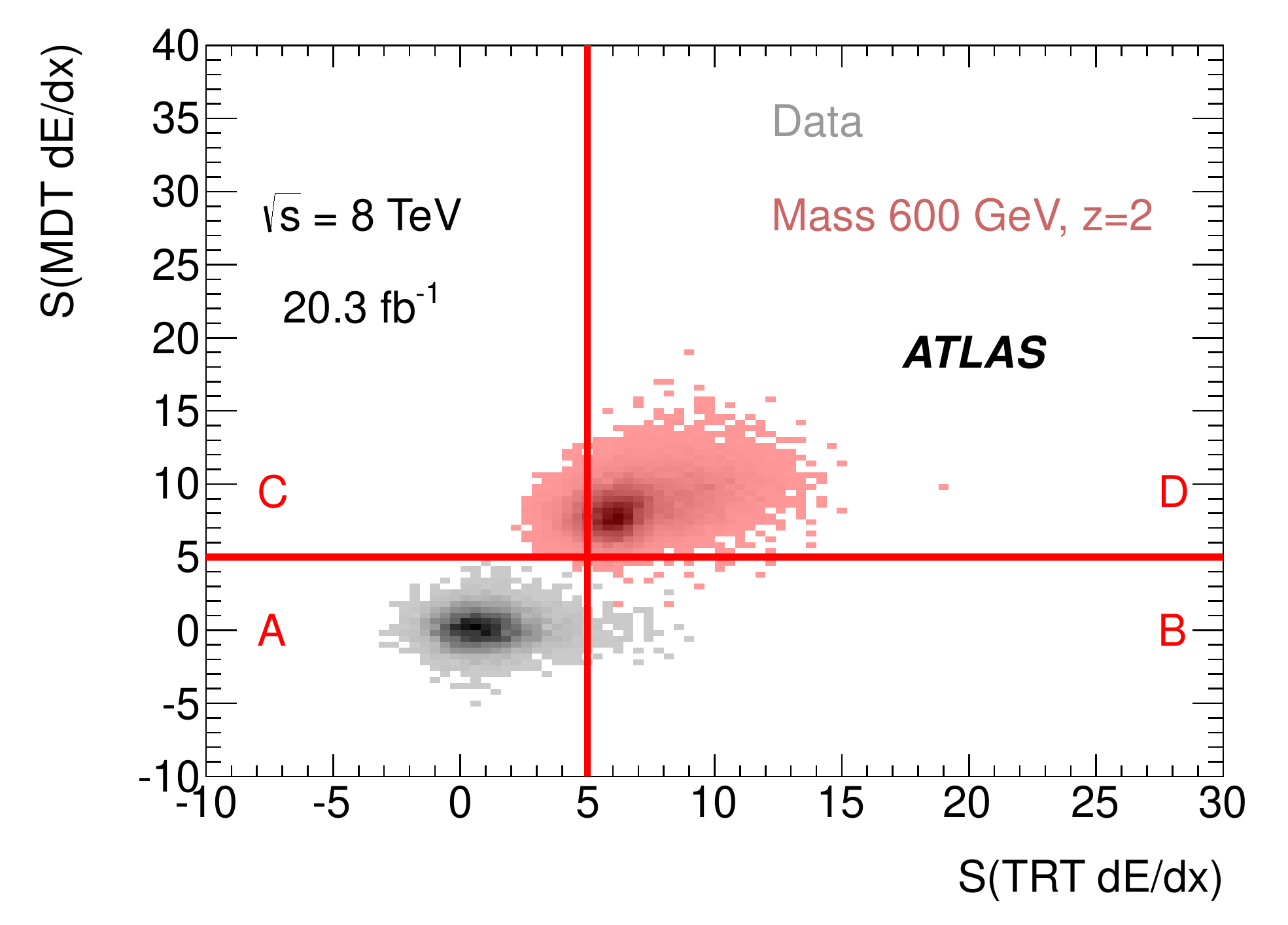}
    \includegraphics[width=0.45\textwidth]{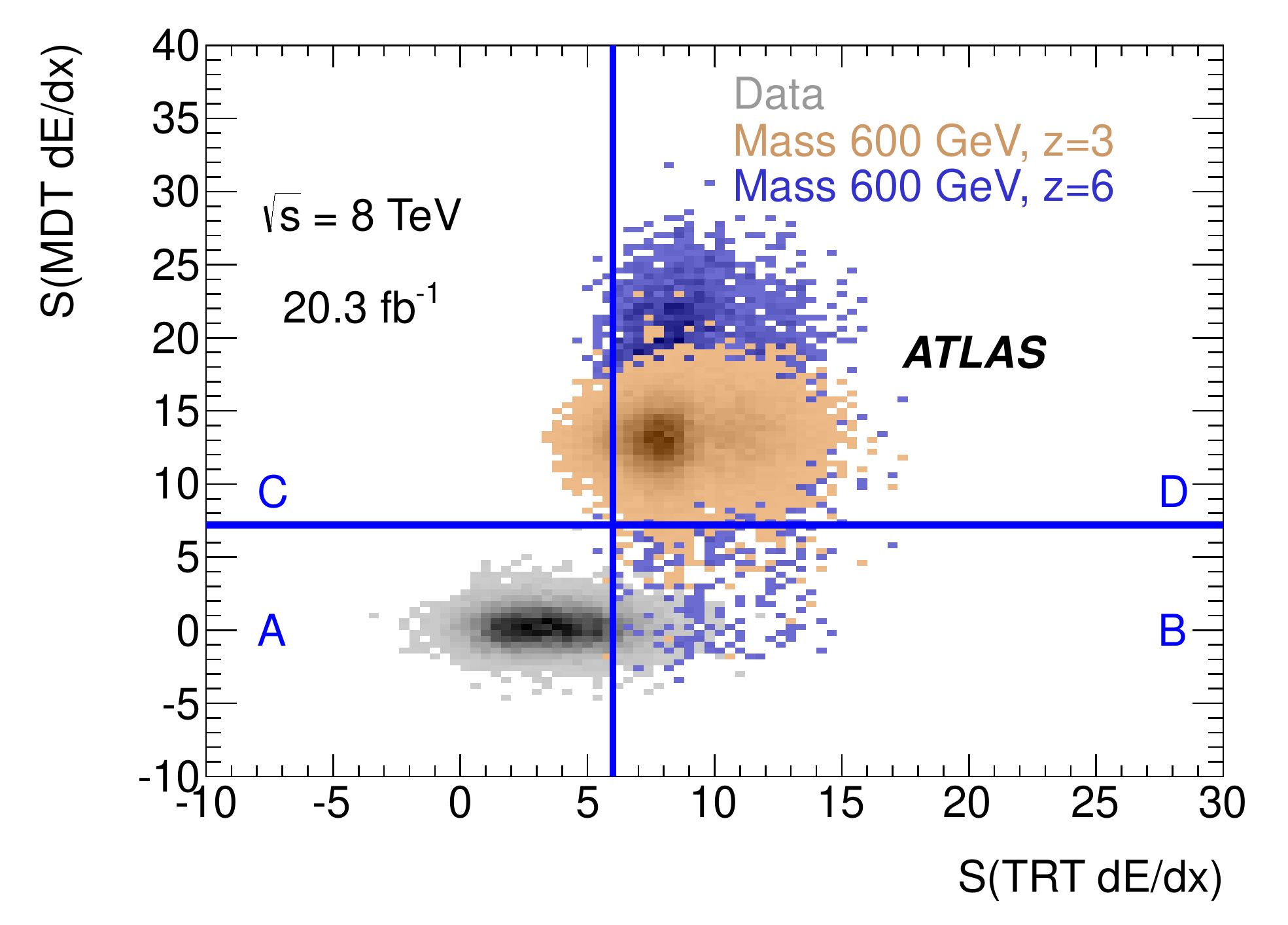}
  \end{center}
  \caption{$S$(MDT \dedx{}) versus $S$(TRT \dedx{}) after 
    the ${z=2}$ (left) or ${z \geq 3}$ (right) tight selection. The distributions of the
    data and the simulated signal samples (here for a mass of \unit[600]{\GeV{}}) are shown. 
    The meaning of the A, B, C and D regions is discussed in the text.
}
  \label{fig:abcplaneq23}
\end{figure}

Two-dimensional distributions of $S$(MDT \dedx{}) versus $S$(TRT \dedx{}) 
are shown for data and simulated signal events 
in Fig.~\ref{fig:abcplaneq23} for candidates
passing the tight selection as ${z = 2}$ (left) and ${z \geq 3}$ (right), 
and also satisfying all previous selection criteria. 
As seen, the sub-detector system signatures are
different for the two preselected samples, and thus the final signal
regions are chosen differently. They are defined by
$S$(MDT \dedx{})~$> \MDTcutQTwo$ and $S$(TRT \dedx{})~$> \TRTcutQTwo$ for 
candidates selected as ${z = 2}$ and by $S$(MDT \dedx{})~$> \MDTcutQThree$ and
$S$(TRT \dedx{})~$> \TRTcutQThree$ for candidates selected as ${z \geq 3}$. 
The selection was optimised using only simulated 
samples and $Z\to\mu\mu$ data control samples without examining the signal region in the data.

A full summary of the analysis selections is presented in Table~\ref{tab:SelectionSummary_v2}.

\begin{table}[htbp]
\caption{ 
Summary of event selection requirements for the event selections 
based on the single-muon trigger and the \etmiss{} trigger.
}
\begin{center}
\begin{tabular}{|c|c||c|c|c|c|}
\hline
\multicolumn{2}{|c||}{} & Trigger and event & Candidate track & Tight and final & Tight and final \\
\multicolumn{2}{|c||}{} & selection	& selection & selections ($z=2$) & selections ($z\geq3$) \\
\hline
\multirow{22}{*}{\rot{Requirements}} & \multirow{11}{*}{\rot{Single-muon trigger case}} & & Any muon with: & & \\
 & & & & & \\
 & & & \multicolumn{1}{l|}{$N_{\textrm{MDT hits}} \geq 7$} & & \\
 & &  $\geq1$ trigger tight muon & \multicolumn{1}{l|}{$\pT/z > \unit[40]{\GeV{}}$} & & \\
 & & with $\pT/z > \unit[36]{\GeV{}}$ & \multicolumn{1}{l|}{$|\eta| < 2.0$} & & \\
 & & & \multicolumn{1}{l|}{$N_{\textrm{SCT hits}} \geq 6$} & & \\
 & & $\geq1$ reconstructed muon & \multicolumn{1}{l|}{$N_{\textrm{TRT hits}} \geq 10$} & & \\
 & & with $\pT/z > \unit[75]{\GeV{}}$& \multicolumn{1}{l|}{$|d_{0}| < \unit[1.5]{mm}$} & Event passing & Event passing \\
 & & & \multicolumn{1}{l|}{$|z_{0}\sin\theta| < \unit[1.5]{mm}$} & preselection & preselection \\
 & & & \multicolumn{1}{l|}{no other tracks} & having a & having a \\
 & & & \multicolumn{1}{r|}{within $\dR < 0.01$} & muon with: & muon with: \\
\cline{2-4}
 & \multirow{11}{*}{\rot{\etmiss{} trigger case}} & & Any muon with: & & \\
 & & & & \multicolumn{1}{ c| }{$S$(pixel \dedx{}) $> 17$} & \multicolumn{1}{ c| }{$f^{\textrm{HT}} > 0.45$}\\
 & & & \multicolumn{1}{l|}{$N_{\textrm{MDT hits}} \geq 7$} & \multicolumn{1}{ c| }{$S$(MDT \dedx{}) $>5$} & \multicolumn{1}{ c| }{$S$(MDT \dedx{}) $>7.2$}\\
 & & & \multicolumn{1}{l|}{$\pT/z > \unit[30]{\GeV{}}$} & \multicolumn{1}{ c| }{$S$(TRT \dedx{}) $>5$} & \multicolumn{1}{ c| }{$S$(TRT \dedx{}) $>6$} \\
 & & trigger $\etmiss{} > \unit[80]{\GeV{}}$ & \multicolumn{1}{l|}{$|\eta| < 2.0$} & & \\
 & & & \multicolumn{1}{l|}{$N_{\textrm{SCT hits}} \geq 6$} & & \\
 & & $\geq1$ reconstructed muon & \multicolumn{1}{l|}{$N_{\textrm{TRT hits}} \geq 10$} & & \\
 & & with $\pT/z > \unit[60]{\GeV{}}$ & \multicolumn{1}{l|}{$|d_{0}| < \unit[1.5]{mm}$} & & \\
 & & & \multicolumn{1}{l|}{$|z_{0}\sin\theta| < \unit[1.5]{mm}$} & & \\
 & & & \multicolumn{1}{l|}{no other tracks} & & \\
 & & & \multicolumn{1}{r|}{within $\dR < 0.01$} & & \\
\hline
\end{tabular}
\end{center}
\label{tab:SelectionSummary_v2}
\end{table}

\section{Background estimation}
\label{sec:bcgEstimation}

The background contribution to the signal region is estimated using a 
method which
employs sidebands of the two discriminating variables. 
In this method, the plane of $S$(TRT \dedx{}) and $S$(MDT \dedx{}) 
is divided into regions A, B, C and D using the final selection cuts 
as shown in Fig.~\ref{fig:abcplaneq23}.
Region D is defined as the signal region, with regions A, B and C
as control regions. The expected number of candidate events from background in data in 
region D, $\ndataexp^{\textrm{D}}$, is estimated from the number of observed events in data in
region B after tight selection, $\ndataobs^{\textrm{B}}$, and the probability, $f$, to find
a particle with $S$(MDT \dedx{}) $>\MDTcutQTwo$ ($\MDTcutQThree$) before
tight selection for the ${z=2}$ (${z \geq 3}$) search case:

\begin{equation}
\ndataexp^{\textrm{D}} = \ndataobs^{\textrm{B}}\times f.
\label{eq:modifiedABCD}
\end{equation}

\begin{figure}[h!]
  \begin{center}
    \includegraphics[width=0.6\textwidth]{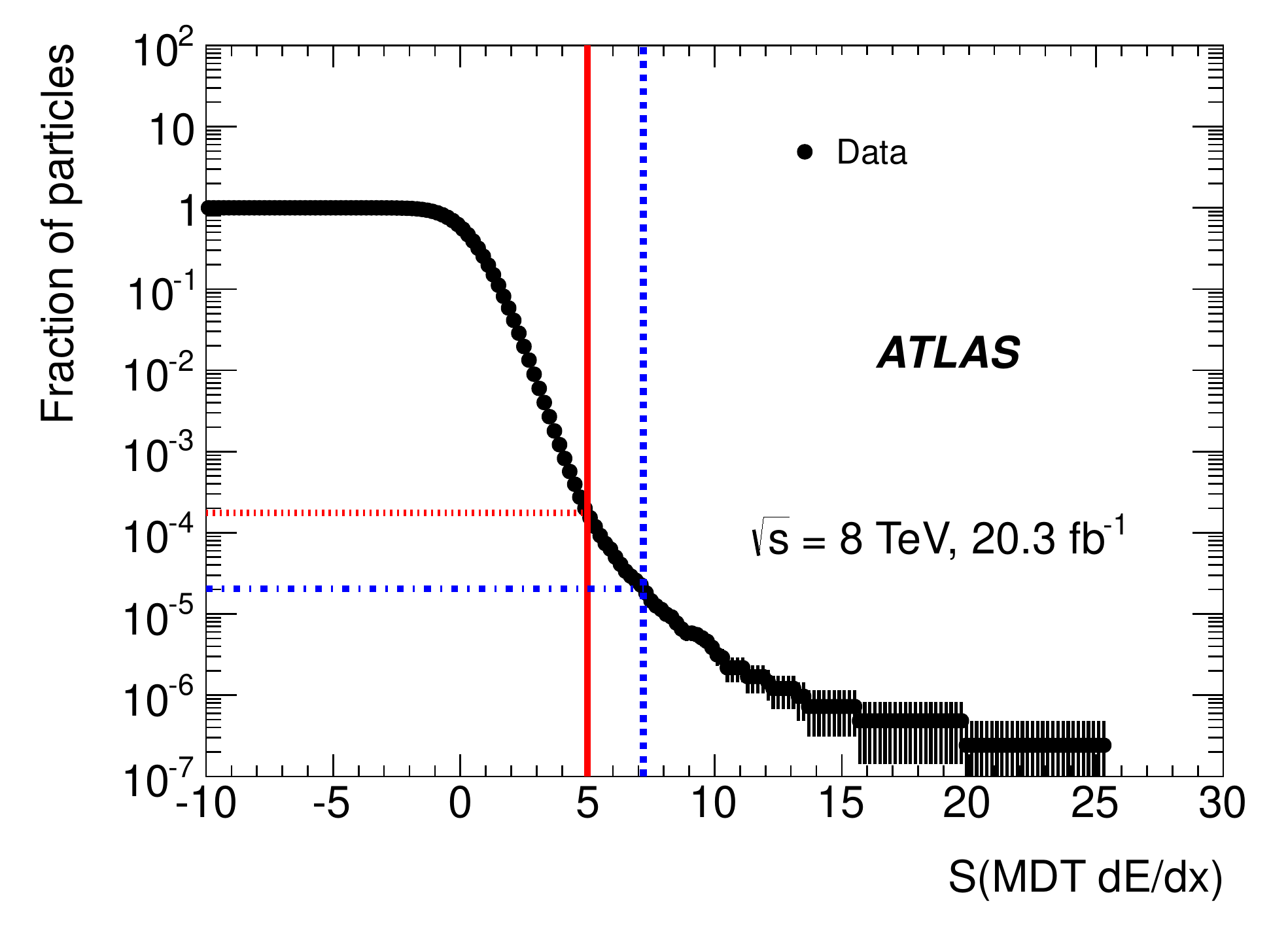}
  \end{center}
  \caption{
Cumulative (from above) $S$(MDT \dedx{}) distribution before tight selection used to
calculate the probability $f$ to find a muon above a certain $S$(MDT \dedx{}) value. 
Indicated in red and blue are the probabilities for $S$(MDT \dedx{}) 
to exceed the values $5$ and $7.2$, respectively. 
}
  \label{fig:MDTProb}
\end{figure}

The probability $f$ to find a particle above some $S$(MDT \dedx{}) value before tight selection 
is derived from the cumulative $S$(MDT \dedx{}) distribution for preselected candidates in 
data shown in Fig.~\ref{fig:MDTProb}.
Although there are no limitations on the $S$(TRT \dedx{}) values of these 
particles, any possible signal contamination in this distribution is negligible.

This method relies on the fact that $S$(MDT \dedx{}) is not correlated with 
the tight selection quantities, 
$S$(pixel \dedx{}), $f^{\textrm{HT}}$ or with $S$(TRT \dedx{}).
A check was performed to demonstrate the absence of such correlations: 
the distributions of $S$(pixel \dedx{}), $f^{\textrm{HT}}$ and $S$(TRT \dedx{}) for 
muons with low $S$(MDT \dedx{}) values were compared with those for muons with 
high $S$(MDT \dedx{}) values. Excellent agreement between the two cases shows that 
there are no correlations between ionisation estimators in different ATLAS sub-detectors 
for background.

Table~\ref{tab:yields_bkg} gives numbers of observed events with particles 
in the B and D regions, as well as the
probabilities to find a particle above certain $S$(MDT \dedx{}) values before tight selection. 
The expected
numbers of background events are given in the last column. They amount to \backgroundEstQTwo{} in 
the signal region for the ${z = 2}$ selection
and \backgroundEstQThree{} for the ${z \geq 3}$ selection, where the quoted uncertainties are
statistical. Systematic uncertainties on the background estimate are discussed 
in Sect.~\ref{sec:systematics}.

\begin{table}[htbp]
\caption{The observed event yield in data in the B region, the probability $f$ to find 
  a particle above the respective $S$(MDT \dedx{}) value before tight selection and the expected background yield 
  in the signal region D with its statistical uncertainty. 
  The last column shows the observed event yield in the D region.}
\begin{center}
\begin{tabular}{|c||c c c|c|}
\hline
\rule{0pt}{3ex} & $\ndataobs^{\textrm{B}}$ & $f$ & $\ndataexp^{\textrm{D}}$ & $\ndataobs^{\textrm{D}}$ \\
\hline
$z=2$  		& $76$   & $1.8 \times 10^{-4}$   & \multicolumn{1}{r|}{\backgroundEstQTwo{}}   & $0$ \\
$z \geq 3$  	& $1251$ & $2.1 \times 10^{-5} $  & \multicolumn{1}{r|}{\backgroundEstQThree{}} & $0$ \\
\hline
\end{tabular}
\end{center}
\label{tab:yields_bkg}
\end{table}

\section{Signal efficiency}
\label{sec:efficiency}

\begin{figure}[h!]
  \begin{center}
    \includegraphics[width=0.6\textwidth]{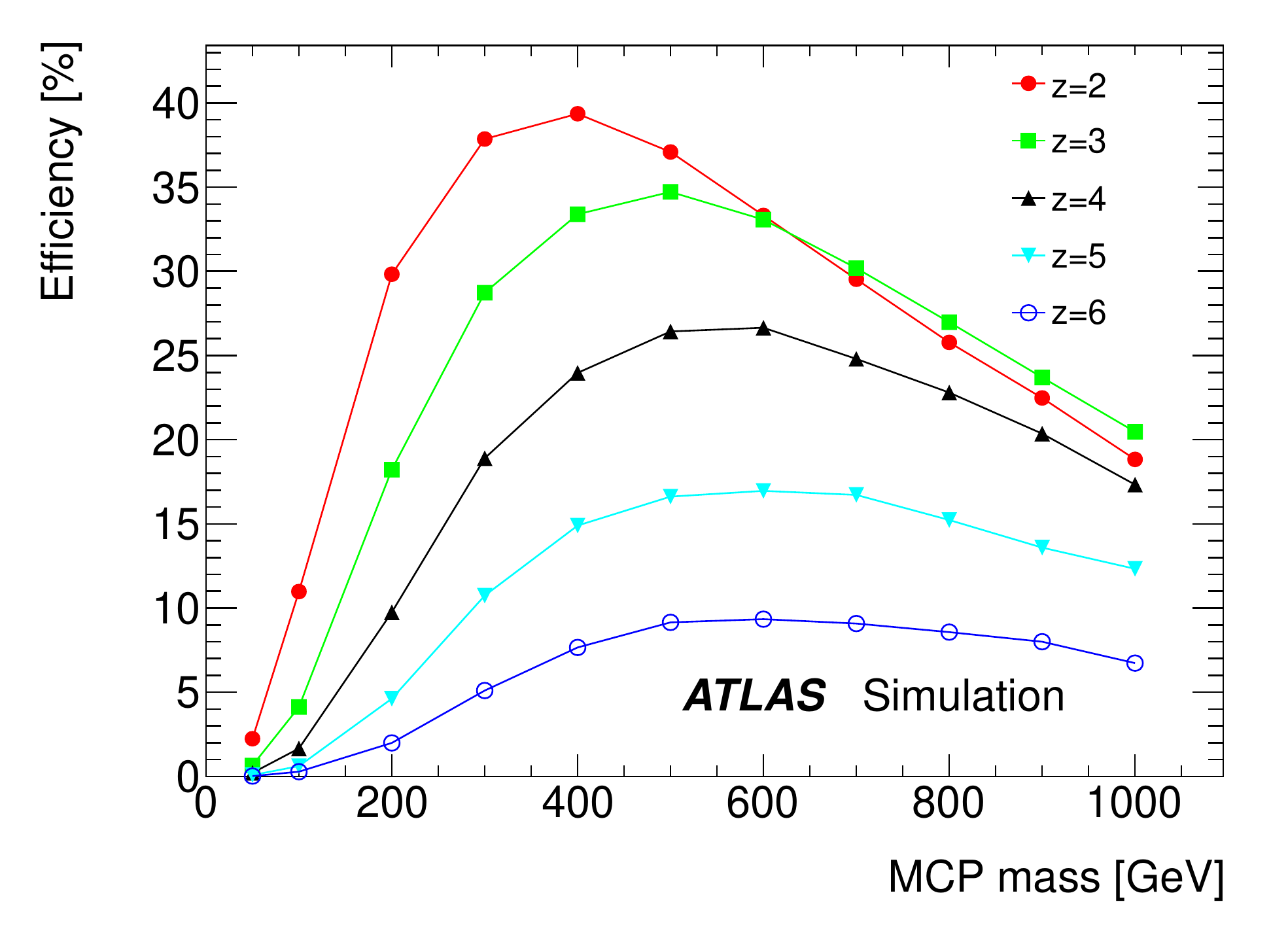}
  \end{center}
  \caption{The signal efficiencies for different MCP masses and charges for the DY production model.}
  \label{fig:EfficiencyTrends}
\end{figure}

The cross-section is given by

\begin{equation}
\sigma = \frac{\ndataobs^{\textrm{D}}-\ndataexp^{\textrm{D}}}{\mathcal{L}\times\varepsilon},
\end{equation}
where $\mathcal{L}$ is the integrated luminosity of the analysed data and
the numerator is the number of candidate events above the expected background.
The signal efficiency, $\varepsilon$, includes trigger, reconstruction and selection efficiencies.
The signal efficiency, as estimated from simulation, is shown in Fig.~\ref{fig:EfficiencyTrends} for each signal sample.

Several factors contribute to the efficiency dependence on mass and charge. 
For low masses, the minimum ${\pT / z}$ requirements are the 
main source of efficiency loss. 
At higher masses, the requirement to
reach the MS with a $\beta$ which satisfies the trigger timing window 
is the primary reason for the reduction in efficiency.
Also, high ionisation loss makes particles slow down: they may not 
fit the trigger timing window or may lose all their kinetic energy before reaching the MS.
The charge dependence of the efficiency results from the higher ionisation loss and 
the higher effective $\pT$ selection, which are augmented 
by factors $z^2$ and $z$, respectively. 
For MCPs that do not reach the MS, the \etmiss{} would be larger for heavier MCPs 
and therefore more likely to fire the \etmiss{} trigger, 
although the probability for such events to satisfy all selection criteria is smaller since 
only one candidate of an MCP pair is reconstructed in the MS.

The fraction of signal events satisfying cumulative selection
requirements is given in Table~\ref{tab:EfficiencyCutflow} for several
examples. 
 
\begin{table*}[ht]
\caption{Fractions of signal events (in \%) with at least one MCP, which satisfy the given requirements. The uncertainties quoted are statistical.}
\centering
\begin{tabular}{|c||c|c|c|c|c|c|}
\hline
Signal benchmark point	        & Trigger 	& Preselection	& Tight selection	& Final selection	\\
\hline
\hline
 $m$=\unit[100]{\GeV{}}, $z=2$ & $13.7\pm0.2$  & $12.8\pm0.2$ & $12.6\pm0.2$ & $11.0\pm0.2$ \\
 $m$=\unit[500]{\GeV{}}, $z=2$ & $62.8\pm0.4$  & $42.9\pm0.3$ & $39.4\pm0.3$ & $37.1\pm0.3$ \\
 $m$=\unit[900]{\GeV{}}, $z=2$ & $35.2\pm0.4$  & $26.6\pm0.3$ & $24.4\pm0.3$ & $22.5\pm0.3$ \\
\hline
 $m$=\unit[100]{\GeV{}}, $z=4$ & $\enspace 2.01\pm0.09$ & $\enspace 1.74\pm0.08$ & $\enspace 1.71\pm0.08$ & $\enspace 1.66\pm0.08$ \\
 $m$=\unit[500]{\GeV{}}, $z=4$ & $32.5\pm0.3$  & $28.7\pm0.3$ & $28.2\pm0.3$ 	     & $26.4\pm0.3$ \\
 $m$=\unit[900]{\GeV{}}, $z=4$ & $29.7\pm0.4$  & $22.4\pm0.3$ & $21.8\pm0.3$ 	     & $20.4\pm0.3$ \\
\hline
 $m$=\unit[50]{\GeV{}}, $z=6$  & $\enspace 0.04\pm0.02$	& $\enspace 0.03\pm0.02$  & $\enspace 0.03\pm0.02$     & $\enspace 0.02\pm0.01$ \\ 
 $m$=\unit[100]{\GeV{}}, $z=6$ & $\enspace 0.58\pm0.08$ & $\enspace 0.35\pm0.05$  & $\enspace 0.32\pm0.04$ & $\enspace 0.28\pm0.04$ \\
 $m$=\unit[500]{\GeV{}}, $z=6$ & $16.2\pm0.4$  & $10.3\pm0.3$ 	& $10.0\pm0.2$ 		 & $\enspace 9.2\pm0.2$ \\
 $m$=\unit[900]{\GeV{}}, $z=6$ & $17.4\pm0.6$  & $\enspace 9.5\pm0.4$ 	& $\enspace 9.0\pm0.3$ 		 & $\enspace 8.0\pm0.2$ \\
\hline
\end{tabular}
\label{tab:EfficiencyCutflow}
\end{table*}

\section{Systematic uncertainties}
\label{sec:systematics}

Systematic uncertainties of the analysis comprise the uncertainty on the background estimate, 
on the signal selection efficiency, on the luminosity, and the one due to 
the size of the Monte Carlo samples used.

\subsection{Background estimation uncertainty}
\label{sec:beu}

A difference is assessed between the current method and an alternate method 
(ABCD method, as used in Ref.~\cite{Qballs2011}) where the number of expected 
events from background is calculated 
from the numbers of observed events in the three control regions according to

\begin{equation}
\ndataexp^{\textrm{D}} = \frac{\ndataobs^{\textrm{B}}\times \ndataobs^{\textrm{C}}}{\ndataobs^{\textrm{A}}}.
\label{eq:regularABCD}
\end{equation}
Both methods use the same underlying idea, that the background estimate is 
proportional to the number of observed events in the region B, $\ndataobs^{\textrm{B}}$.
However, the methods to derive the proportionality constant are 
different, cf. Eq.~(\ref{eq:modifiedABCD}) and Eq.~(\ref{eq:regularABCD}).

Since the ABCD method 
gives a large statistical uncertainty
in the case of zero events in one of the control regions, the cuts on $S$(MDT \dedx{}) were 
loosened from $\MDTcutQTwo$ or $\MDTcutQThree$ down to $3$ for both the ${z=2}$ and
${z \geq 3}$ selections to minimise this uncertainty, and the numbers of events expected from the background
were re-estimated using the two aforementioned methods. 
The background estimates from the two methods were found to differ by 
about $25\%$ for both the ${z=2}$ and ${z\geq3}$ cases, corresponding for 
both to a statistical significance of less than two sigma. Hence, a $25\%$ 
systematic uncertainty on the background estimate was assigned for both 
the ${z=2}$ and ${z\geq3}$ cases.

\subsection{Trigger efficiency uncertainty}

The uncertainty on the muon trigger efficiency has two sources: 
a global uncertainty on the muon trigger 
efficiency of 1\%~\cite{Aad:2014sca} 
and a $\beta$-dependent uncertainty.
The $\beta$-dependent part originates from uncertainties on 
the modelling of the muon trigger timing for particles with $\beta < 1$. 
In order to improve the description of the trigger simulation, parameterised corrections were applied.
To assess the uncertainties, the parameters of these corrections
were varied. The $\beta$ value of particles was varied 
between the true generated value 
and the one reconstructed in the MS from the known mass and measured momentum.\footnote{The 
relation between a particle’s $\beta$, $\pT$ and mass, $m$, is given by
$\beta = \frac{\pT/\sin\theta}{\sqrt{\left(\pT/\sin\theta\right)^2+m^2}}$.} 
The time interval needed for a signal particle to reach the
RPC trigger planes was varied by the root-mean-square width of the timing distribution for 
muons measured in the full $Z\to\mu\mu$ sample in data.
The combination of these effects 
ranges from $0.4\%$ to $13\%$. 
The timing in the TGC for data and simulation is in 
good agreement, and the systematic uncertainty for the TGC timing correction is negligible.

The uncertainty on the \etmiss{} trigger efficiency consists of two parts: a global $5\%$ uncertainty due 
to a difference between triggering in data and simulation~\cite{MetTriggerGlobalUncNote}
especially in the turn-on region, 
and $8.5\%$ uncertainty due to the fact that the \etmiss{} trigger efficiency 
depends on the amount of initial- and final-state radiation~\cite{SMP2012}, affecting 
the number of signal events which pass the \etmiss{} trigger requirements.
Varying the amount of radiation in the MC, the number of jets in an event 
was altered, and the relative difference of the \etmiss{} trigger efficiency 
was taken as a systematic uncertainty.

\subsection{Uncertainties due to selection}

The uncertainty on the selection efficiency is evaluated by varying 
the requirement values used in the analysis. Several reasons motivate these variations. 
For example, the uncertainty on the amount of material in front of the MS, 
which is found to be about $1\%$~\cite{Aad:2014rra},  
propagates into an uncertainty on the selection efficiency due to the slowing down of particles, 
and its effect is covered by the effect of varying the \pT requirement.
The following variations of the nominal requirements are studied:
$\pT$ value by $\pm3$\% because of an uncertainty on the track $\pT$ measurements 
and the uncertainty on the amount of material; 
$f^{\textrm{HT}}$ value by $\pm25$\% due to pile-up dependence,
$S$(pixel \dedx{}) by $\pm10$\%, $S$(TRT \dedx{}) by $\pm5$\% and $S$(MDT \dedx{}) by $\pm15$\%
because of the observed disagreement of the mean and root-mean-square width of these 
distributions in the $Z\to\mu\mu$ events in data and simulation, as well as of any potential 
mismodelling of these ionisation estimators.

For all other variables the variations have no observable effect in any
of the signal samples. The total systematic uncertainties on the efficiency 
arising from these variations range between $1\%$ and $17\%$, where the larger 
uncertainty corresponds to lower-mass signal samples. 
This uncertainty is dominated by the effect of the \pT requirement 
variation, which the lightest MCPs are most sensitive to.

The uncertainties due to the choice of parton distribution functions and due to 
higher orders corrections propagate into a small uncertainty on the selection 
efficiency, which lies well within its statistical uncertainty.

\subsection{Summary of systematic uncertainties}

The contributions from the separate sources of systematic uncertainty on the signal efficiency 
are shown in Table~\ref{tab:UncertaintyIngredients} for 
several charges and mass points. The uncertainties on the luminosity 
and due to limited Monte Carlo samples size are also shown.
Since the expected number of events from background is close to zero, 
the $25\%$ uncertainty on this number has a very small effect on the 
calculation of the upper limit on the cross-section. 
Thus, the trigger and selection efficiencies are the main sources of uncertainty.
An additional statistical uncertainty to take into account the limited size 
of the Monte Carlo samples is added.
The samples with a mass of \unit[50]{\GeV} and charge numbers $z=5$, $z=6$ were produced with a selection
at the generator level requiring ${\pT/z > }$ \unit[20]{\GeV} in order to decrease 
this uncertainty.
Generally, it is about $3\%$, although it makes a
significant contribution (up to $60\%$) for high-charge and low-mass samples.

The uncertainty on the integrated luminosity is $2.8\%$. It is derived, following the 
same methodology as that detailed in Ref.~\cite{lumiRef}, from a 
calibration of the luminosity scale derived from beam-separation scans performed in November 2012.

\begin{table*}[ht]
\caption{Overview of separate contributions (in \%) to the systematic uncertainty on the signal. The total uncertainty is 
   given by the quadratic sum of the individual uncertainties.}
\centering
\begin{tabular}{|c||c|c|c|c||c|}
\hline
 		& 		& 		   & Limited	 & 	         & \\
Signal benchmark& Trigger 	& Selection 	   & Monte Carlo & Luminosity  	 & Total uncertainty \\
point		& efficiency	& efficiency	   & samples size&		 & \\
\hline
\hline
 $m$=\unit[100]{\GeV{}}, $z=2$ & $6.1$	& $11$ 	& $1.8$ 	& \multirow{10}{*}{$2.8$}& $13$ \\
 $m$=\unit[500]{\GeV{}}, $z=2$ & $8.9$	& $4.7$	& $0.8$ 	&			& $11$ \\
 $m$=\unit[900]{\GeV{}}, $z=2$ & $9.7$	& $1.8$	& $1.2$ 	&			& $10$ \\
\cline{1-4}\cline{6-6}
 $m$=\unit[100]{\GeV{}}, $z=4$ & $3.9$	& $8.5$	& $5.1$ 	& 			& $11$ \\
 $m$=\unit[500]{\GeV{}}, $z=4$ & $9.7$	& $2.9$	& $1.1$ 	&			& $11$ \\
 $m$=\unit[900]{\GeV{}}, $z=4$ & $8.9$	& $1.3$	& $1.3$ 	&			& $9.5$ \\
\cline{1-4}\cline{6-6}
 $m$=\unit[50]{\GeV{}}, $z=6$  & $4.0$	& $13$	& $60$		& 			& $61$ \\
 $m$=\unit[100]{\GeV{}}, $z=6$ & $4.0$	& $17$	& $13$  	&			& $22$ \\
 $m$=\unit[500]{\GeV{}}, $z=6$ & $11$	& $4.1$	& $2.0$ 	&			& $12$ \\
 $m$=\unit[900]{\GeV{}}, $z=6$ & $10$	& $3.0$	& $2.2$ 	&			& $11$ \\
\hline
\end{tabular}
\label{tab:UncertaintyIngredients}
\end{table*}

\section{Results}
\label{sec:limits}

No signal candidate events are found for either the ${z=2}$ or the ${z \geq 3}$ selections. 
The results are consistent with the expectation of \backgroundEstQTwoStatSys{} and
\backgroundEstQThreeStatSys{} background events, 
respectively. Since the number of signal events expected from background 
is very small and consistent with the observation of zero candidate events, 
observed and expected limits are virtually identical. 
The limits are computed with
MCLimit~\cite{Junk:1999kv}. It uses the CL$_s$  method~\cite{CLsMethod} to discriminate 
between the background-only hypothesis and  
the signal-plus-background hypothesis, 
and determines exclusion limits for various MCP scenarios. The signal selection 
efficiency, luminosity, their uncertainties and number of observed events 
are taken as input for pseudo-experiments, 
resulting in an observed limit at $95\%$ confidence level (CL). 

\begin{figure}[h!]
 \centering
 \includegraphics[width=0.6\textwidth]{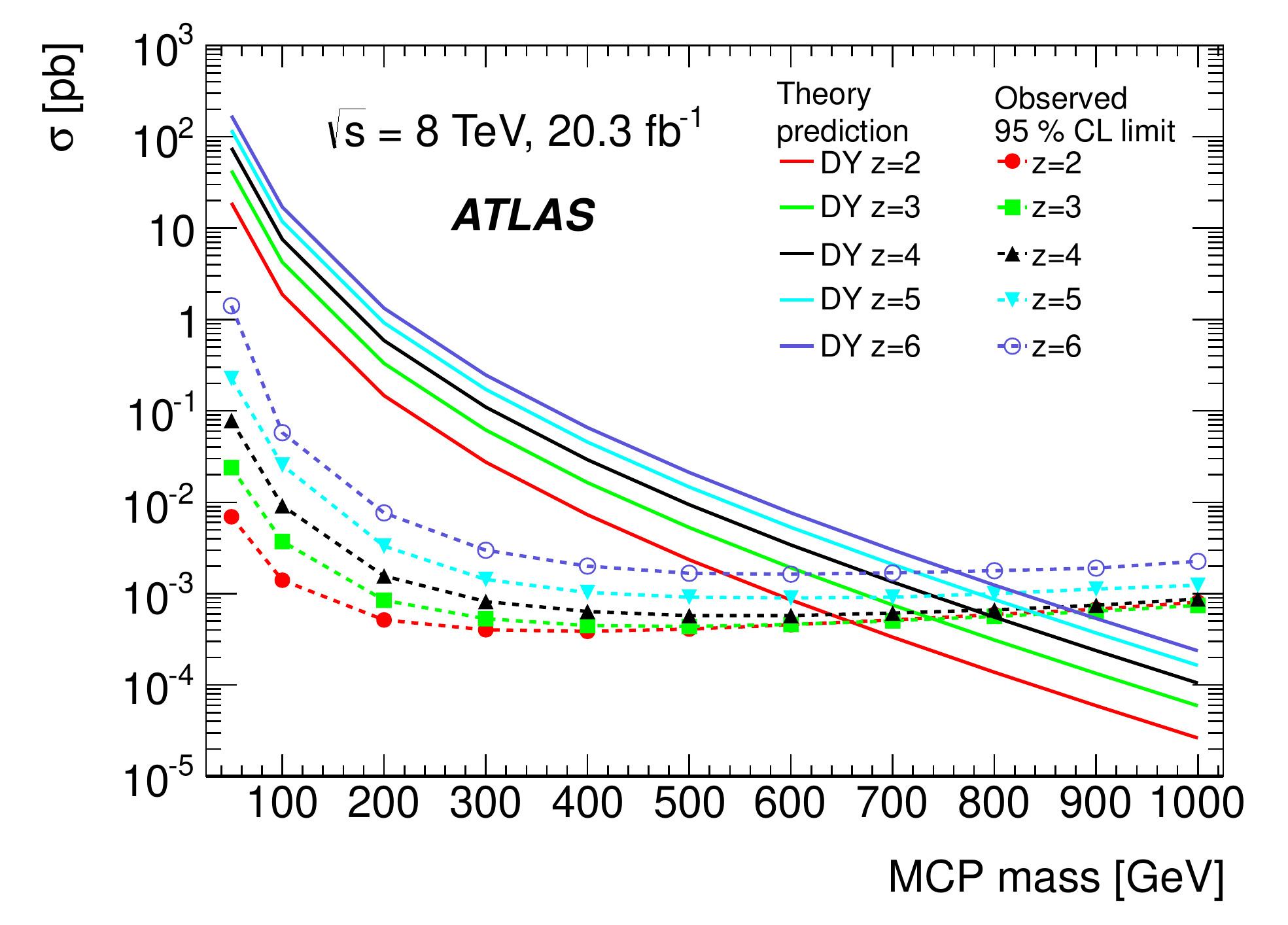}
 \caption{Observed $95\%$ CL cross-section upper limits and theoretical cross-sections as functions of the MCP's mass for 
   values of $z$ between $2$ and $6$.}
 \label{fig:limtrack}
\end{figure}

The measurement excludes the DY model of MCP pair-production over wide ranges of tested masses. 
Figure~\ref{fig:limtrack} shows the observed $95\%$ CL cross-section limits
as a function of mass for the five different charges. 
At the lowest mass values the cross-section limit ranges 
from \unit[7]{fb} for $z=2$ to \unit[1.4]{pb} for $z=6$. The most stringent cross-section 
limits are obtained for masses of about \unit[400]{\GeV{}} and range from $0.4$ to \unit[1.6]{fb}.
In addition, the theoretical cross-section is shown for the
simplified Drell--Yan model. 
The uncertainty on the theoretical cross-section is due to the parton distribution functions 
choice and is estimated to be $5\%$.
For this model, the cross-section limits can be transformed into 
mass exclusion regions from \unit[50]{\GeV{}} up to limits 
of $660$, $740$, $780$, $785$, and \unit[760]{\GeV{}} for 
charge numbers ${z = 2}$, $3$, $4$, $5$, and $6$, respectively. 
Mass limits are obtained from the intersection of the observed limits and the 
central values of the theoretical cross-section.
This result is similar to that 
obtained by the CMS collaboration~\cite{QballsCms20112012} and extends the excluded region 
approximately \unit[300]{\GeV{}} further than in the previous ATLAS search~\cite{Qballs2011}.

\section{Conclusion}
\label{sec:conclusion}

This article reports on a search for long-lived multi-charged particles 
produced in proton--proton collisions with the ATLAS detector at the LHC. 
The search uses a data sample with a center-of-mass-energy of $\sqrt{s}=$ \unit[8]{\TeV} 
and an integrated luminosity of \intlumi{}. Particles with electric 
charges from ${|q|=2e}$ to ${|q|=6e}$ penetrating the full ATLAS detector 
and producing anomalously high ionisation signals in multiple detector 
elements are searched for. Less than one background event is expected and 
no events are observed. Upper limits are derived on the production cross-sections 
and are interpreted as mass exclusion limits for a Drell-Yan 
production model from \unit[50]{\GeV{}} up to $660$, $740$, $780$, $785$, 
and \unit[760]{\GeV{}} for charges ${|q| = 2e}$, $3e$, $4e$, $5e$, and $6e$, respectively.

\section*{Acknowledgements}


We thank CERN for the very successful operation of the LHC, as well as the
support staff from our institutions without whom ATLAS could not be
operated efficiently.

We acknowledge the support of ANPCyT, Argentina; YerPhI, Armenia; ARC,
Australia; BMWFW and FWF, Austria; ANAS, Azerbaijan; SSTC, Belarus; CNPq and FAPESP,
Brazil; NSERC, NRC and CFI, Canada; CERN; CONICYT, Chile; CAS, MOST and NSFC,
China; COLCIENCIAS, Colombia; MSMT CR, MPO CR and VSC CR, Czech Republic;
DNRF, DNSRC and Lundbeck Foundation, Denmark; EPLANET, ERC and NSRF, European Union;
IN2P3-CNRS, CEA-DSM/IRFU, France; GNSF, Georgia; BMBF, DFG, HGF, MPG and AvH
Foundation, Germany; GSRT and NSRF, Greece; RGC, Hong Kong SAR, China; ISF, MINERVA, GIF, I-CORE and Benoziyo Center, Israel; INFN, Italy; MEXT and JSPS, Japan; CNRST, Morocco; FOM and NWO, Netherlands; BRF and RCN, Norway; MNiSW and NCN, Poland; GRICES and FCT, Portugal; MNE/IFA, Romania; MES of Russia and NRC KI, Russian Federation; JINR; MSTD,
Serbia; MSSR, Slovakia; ARRS and MIZ\v{S}, Slovenia; DST/NRF, South Africa;
MINECO, Spain; SRC and Wallenberg Foundation, Sweden; SER, SNSF and Cantons of
Bern and Geneva, Switzerland; NSC, Taiwan; TAEK, Turkey; STFC, the Royal
Society and Leverhulme Trust, United Kingdom; DOE and NSF, United States of
America.

The crucial computing support from all WLCG partners is acknowledged
gratefully, in particular from CERN and the ATLAS Tier-1 facilities at
TRIUMF (Canada), NDGF (Denmark, Norway, Sweden), CC-IN2P3 (France),
KIT/GridKA (Germany), INFN-CNAF (Italy), NL-T1 (Netherlands), PIC (Spain),
ASGC (Taiwan), RAL (UK) and BNL (USA) and in the Tier-2 facilities
worldwide.

\providecommand{\href}[2]{#2}\begingroup\raggedright

\endgroup

\newpage 
\begin{flushleft}
{\Large The ATLAS Collaboration}

\bigskip

G.~Aad$^{\rm 85}$,
B.~Abbott$^{\rm 113}$,
J.~Abdallah$^{\rm 152}$,
O.~Abdinov$^{\rm 11}$,
R.~Aben$^{\rm 107}$,
M.~Abolins$^{\rm 90}$,
O.S.~AbouZeid$^{\rm 159}$,
H.~Abramowicz$^{\rm 154}$,
H.~Abreu$^{\rm 153}$,
R.~Abreu$^{\rm 30}$
% ...
\bigskip
\\
% ...
$^{11}$ Institute of Physics, Azerbaijan Academy of Sciences, Baku, Azerbaijan\\
% ...
$^{30}$ CERN, Geneva, Switzerland\\
% ...
$^{85}$ CPPM, Aix-Marseille Universit{\'e} and CNRS/IN2P3, Marseille, France\\
% ...
$^{90}$ Department of Physics and Astronomy, Michigan State University, East Lansing MI, United States of America\\
% ...
$^{107}$ Nikhef National Institute for Subatomic Physics and University of Amsterdam, Amsterdam, Netherlands\\
% ...
$^{113}$ Homer L. Dodge Department of Physics and Astronomy, University of Oklahoma, Norman OK, United States of America\\
% ...
$^{152}$ Institute of Physics, Academia Sinica, Taipei, Taiwan\\
$^{153}$ Department of Physics, Technion: Israel Institute of Technology, Haifa, Israel\\
$^{154}$ Raymond and Beverly Sackler School of Physics and Astronomy, Tel Aviv University, Tel Aviv, Israel\\
% ...
$^{159}$ Department of Physics, University of Toronto, Toronto ON, Canada\\
% ...
\end{flushleft}


\end{document}